\documentclass[journal]{IAENGtran}
\ifCLASSINFOpdf
   \usepackage[pdftex]{graphicx}
   \DeclareGraphicsExtensions{.pdf,.jpeg,.png}
\else
   \usepackage[dvips]{graphicx}
   \DeclareGraphicsExtensions{.eps}
\fi

\usepackage[T1]{fontenc}
\usepackage[latin9]{inputenc}
\usepackage{listings}
\usepackage{float}
\usepackage{amsbsy}
\usepackage{amsmath}

\floatstyle{ruled}
\newfloat{algorithm}{tbp}{loa}
\providecommand{\algorithmname}{Algorithm}
\floatname{algorithm}{\protect\algorithmname}

\begin{document}
%
\title{Portfolio Optimization in R}
%
%
%

\author{M. Andrecut
\thanks{Manuscript received June 26, 2013.}
\thanks{M. Andrecut is with the Unlimited Analytics Inc., Calgary, AB, Canada, e-mail: mircea.andrecut@gmail.com.}
}
\maketitle

\pagestyle{empty}
\thispagestyle{empty}

\begin{abstract}
We consider the problem of finding the efficient frontier associated
with the risk-return portfolio optimization model. We derive
the analytical expression of the efficient frontier for a portfolio
of $N$ risky assets, and for the case when a risk-free asset is added
to the model. Also, we provide an R implementation,
and we discuss in detail a numerical example of a portfolio of several
risky common stocks. 
\end{abstract}

\begin{IAENGkeywords}
portfolio optimization, efficient frontier, R.
\end{IAENGkeywords}

%
\IAENGpeerreviewmaketitle

\section{Introduction}
%
%
%
%
\IAENGPARstart{P}{ortfolio} 
optimization is a challenging problem in economic analysis
and risk management, which dates back to the seminal work of Markowitz
[1]. The main assumption is that the return of any financial asset
is described by a random variable, whose expected mean and variance
are assumed to be reliably estimated from historical data. The expected
mean and variance are interpreted as the reward, and respectively
the risk of the investment. The portfolio optimization problem can
be formulated as following: given a set of financial assets, characterized
by their expected mean and their covariances, find the optimal weight
of each asset, such that the overall portfolio provides the smallest
risk for a given overall return [1-5]. Therefore, the problem
reduces to finding the "efficient frontier", which is the set
of all achievable portfolios that offer the highest rate of return
for a given level of risk. Using the quadratic optimization mathematical
framework it can be shown that for each level of risk there is exactly
one achievable portfolio offering the highest rate of return. Here,
we consider the standard risk-return portfolio optimization model,
when both long buying and short selling of a relatively large number
of assets is allowed. We derive the analytical expression of the efficient
frontier for a portfolio of $N$ risky assets, and for the case when
a risk-free asset is added to the model. Also, we provide an R implementation
for both cases, and we discuss in detail a numerical example of a
portfolio of several risky common stocks. 

\section{Assets and portfolios}

A portfolio is an investment made in $N$ assets $A_{n}$, with the
returns $R_{n}$, $n=1,2,...,N$, using some amount of wealth $W$.
Let $W_{n}$ denote the amount invested in the $n$-th asset. Negative
values of $W_{n}$ can be interpreted as short selling. Since the
total wealth is $W$ we have:
\begin{equation}
\sum_{n=1}^{N}W_{n}=W.
\end{equation}
It is convenient to describe the investments in terms of relative
values such that:
\begin{equation}
w_{n}=W_{n}/W,\quad\sum_{n=1}^{N}w_{n}=1,
\end{equation}
and
\begin{equation}
\sum_{n=1}^{N}w_{n}R_{n}=R.
\end{equation}

To characterize the portfolio we consider the expected return:
\begin{equation}
\rho=E(R)=E(\sum_{n=1}^{N}w_{n}R_{n})=\sum_{n=1}^{N}w_{n}r_{n},
\end{equation}
where $r_{n}=E(R_{n})$ is the expected return of each asset, $n=1,2,...,N$.
Also, we use the covariance matrix of the portfolio:
\begin{equation}
\mathbf{S}=\left[\begin{array}{cccc}
s_{11} & s_{12} & \cdots & s_{1N}\\
s_{21} & s_{22} & \cdots & s_{2N}\\
\vdots & \vdots & \ddots & \vdots\\
s_{N1} & s_{N2} & \cdots & s_{NN}
\end{array}\right],
\end{equation}
where
\begin{equation}
s_{ij}=s_{ji}=E((R_{i}-r_{i})(R_{j}-r_{j})),
\end{equation}
in order to quantify the deviation from the expected return, and to
capture the risk of the investment. The variance of the portfolio
is then given by:
\begin{equation}
s^{2}=E(\left|R-\rho\right|^{2})=\sum_{i=1}^{N}\sum_{j=1}^{N}w_{i}w_{j}s_{ij}=\mathbf{w}^{T}\mathbf{S}\mathbf{w},
\end{equation}
where $\mathbf{w}=[w_{1},w_{2},\ldots,w_{N}]^{T}$ is the vector of
weights. 

\section{\textit{N} risky assets}

A portfolio is optimal if for a given expected return $\rho$, the
portfolio has the least variance $s^{2}$. Finding such a portfolio
requires the solution of the following constrained quadratic optimization
problem [6]:
\begin{equation}
\mathbf{w}=\arg\min_{\mathbf{w}}\left\{ \mathbf{w}^{T}\mathbf{S}\mathbf{w}\right\} ,
\end{equation}
subject to:
\begin{itemize}
\item the constant invested wealth constraint (equivalent to Eq. 2)
\end{itemize}
\begin{equation}
\mathbf{w}^{T}\mathbf{u}=\sum_{n=1}^{N}w_{n}=1,
\end{equation}

\begin{itemize}
\item the expected return constraint (equivalent to Eq. 4):
\end{itemize}
\begin{equation}
\mathbf{w}^{T}\mathbf{r}=\sum_{n=1}^{N}w_{n}r_{n}=\rho,
\end{equation}
where $\mathbf{u}=[1,1,\ldots,1]^{T}$ and $\mathbf{r}=[r_{1},r_{2},\ldots,r_{N}]^{T}$. 

This problem can be solved using the method of Lagrange multipliers. 
Let us define the Lagrangian:
\begin{equation}
L(\mathbf{w},\mu_{1},\mu_{2})=\mathbf{w}^{T}\mathbf{S}\mathbf{w}-\mu_{1}(\mathbf{w}^{T}\mathbf{u}-1)-\mu_{2}(\mathbf{w}^{T}\mathbf{r}-\rho),
\end{equation}
where $\mu_{1}$ and $\mu_{2}$ are the Lagrange multipliers. The
critical point of the Lagrangian can be obtained by solving the system
of equations:
\begin{equation}
\nabla_{\mathbf{w}}L(\mathbf{w},\mu_{1},\mu_{2})=2\mathbf{S}\mathbf{w}-\mu_{1}\mathbf{u}-\mu_{2}\mathbf{r}=0,
\end{equation}
\begin{equation}
\frac{\partial L(\mathbf{w},\mu_{1},\mu_{2})}{\partial\mu_{1}}=\mathbf{w}^{T}\mathbf{u}-1=0,
\end{equation}
\begin{equation}
\frac{\partial L(\mathbf{w},\mu_{1},\mu_{2})}{\partial\mu_{2}}=\mathbf{w}^{T}\mathbf{r}-\rho=0.
\end{equation}
From the first equation we have:
\begin{equation}
\mathbf{w}=\frac{1}{2}\mathbf{S^{\mathrm{-1}}}(\mu_{1}\mathbf{u}+\mu_{2}\mathbf{r}),
\end{equation}
and from the next two equations we have:
\begin{equation}
\mathbf{u}^{T}\mathbf{S}^{-1}\mathbf{u}\mu_{1}+\mathbf{r}^{T}\mathbf{S}^{-1}\mathbf{u}\mu_{2}=2,
\end{equation}
\begin{equation}
\mathbf{u}^{T}\mathbf{S}^{-1}\mathbf{r}\mu_{1}+\mathbf{r}^{T}\mathbf{S}^{-1}\mathbf{r}\mu_{2}=2\rho.
\end{equation}
Taking into account that:
\begin{equation}
\mathbf{u}^{T}\mathbf{S}^{-1}\mathbf{r}=\mathbf{r}^{T}\mathbf{S}^{-1}\mathbf{u},
\end{equation}
we can write:
\begin{equation}
\left[\begin{array}{cc}
a_{11} & a_{12}\\
a_{21} & a_{22}
\end{array}\right]\left[\begin{array}{c}
\mu_{1}\\
\mu_{2}
\end{array}\right]=2\left[\begin{array}{c}
1\\
\rho
\end{array}\right],
\end{equation}
where:
\begin{equation}
\mathbf{A}=\left[\begin{array}{cc}
a_{11} & a_{12}\\
a_{21} & a_{22}
\end{array}\right]=\left[\begin{array}{cc}
\mathbf{u}^{T}\mathbf{S}^{-1}\mathbf{u} & \mathbf{r}^{T}\mathbf{S}^{-1}\mathbf{u}\\
\mathbf{r}^{T}\mathbf{S}^{-1}\mathbf{u} & \mathbf{r}^{T}\mathbf{S}^{-1}\mathbf{r}
\end{array}\right].
\end{equation}
This system has a solution if:
\begin{equation}
d=a_{11}a_{22}-a_{12}^{2}\neq0.
\end{equation}
Since $\mathbf{S}$ is a positive definite matrix, the inverse $\mathbf{S}^{-1}$
is also positive definite, which means that $\mathbf{x}^{T}\mathbf{S}^{-1}\mathbf{x}>0$
for any vector $\mathbf{x}\neq0$. Obviously we have $a_{11}>0$ and
$a_{22}>0$, and:
\begin{equation}
0<(a_{12}\mathbf{u}-a_{11}\mathbf{r})^{T}\mathbf{S}^{-1}(a_{12}\mathbf{u}-a_{11}\mathbf{r})=a_{11}d,
\end{equation}
and therefore we also have $d>0$. The Lagrange multipliers are then
given by:
\begin{equation}
\left[\begin{array}{c}
\mu_{1}\\
\mu_{2}
\end{array}\right]=\frac{2}{d}\left[\begin{array}{c}
a_{22}-a_{12}\rho\\
-a_{12}+a_{11}\rho
\end{array}\right],
\end{equation}
and the weights of the optimal portfolio are:
\begin{equation}
\mathbf{w}(\rho)=\mathbf{f}+\rho\mathbf{g},
\end{equation}
where:
\begin{equation}
\mathbf{f}=\frac{1}{d}\mathbf{S}^{-1}\left(a_{22}\mathbf{u}-a_{12}\mathbf{r}\right),
\end{equation}
\begin{equation}
\mathbf{g}=\frac{1}{d}\mathbf{S}^{-1}\left(-a_{12}\mathbf{u}+a_{11}\mathbf{r}\right).
\end{equation}
The portfolio which minimizes the variance for a specified expected
return is called a "frontier portfolio". It follows that all frontier
portfolios $\mathbf{w}(\rho)$ are a linear combination of the two
portfolios $\mathbf{f}$ and $\mathbf{g}$. 

The variance of the frontier portfolio is:
\begin{equation}
\begin{split}
& s^{2}(\rho)=\mathbf{w}^{T}(\rho)\mathbf{S}\mathbf{w}(\rho)\\
&\quad\quad =\rho^{2}\mathbf{g}^{T}\mathbf{S}\mathbf{g}+\rho(\mathbf{g}^{T}\mathbf{S}\mathbf{f}+\mathbf{f}^{T}\mathbf{S}\mathbf{g})+\mathbf{f}^{T}\mathbf{S}\mathbf{f},
\end{split}
\end{equation}
which can be further simplified as:
\begin{equation}
\begin{split}
&
s^{2}(\rho)=\frac{a_{11}}{d}\left(\rho-\frac{a_{12}}{a_{11}}\right)^{2}+\frac{1}{a_{11}}\\
&\quad\quad=\frac{1}{d}(a_{11}\rho^{2}-2a_{12}\rho+a_{22}).
\end{split}
\end{equation}
This equation represents the "efficient frontier", and it represents
a hyperbola in the $(s,\rho)$-plane. From here we obtain the weights
of the minimum variance portfolio: 
\begin{equation}
\mathbf{w}_{MVP}(\rho)=\mathbf{f}+\frac{a_{12}}{a_{11}}\mathbf{g},
\end{equation}
and the corresponding risk-return values:
\begin{equation}
s_{MVP}(\rho)=\sqrt{1/a_{11}},
\end{equation}
\begin{equation}
\rho_{MVP}=a_{12}/a_{11}.
\end{equation}

An important investment preference on the "efficient frontier"
is the portfolio with the maximum Sharpe ratio [1-3]:
\begin{equation}
\xi=\rho/s.
\end{equation}
The Sharpe ratio represents the expected return per unit of risk.
Therefore, the portfolio with maximum Sharpe ratio $\xi$ gives the
highest expected return per unit of risk, and therefore is the most
"risk-efficient" portfolio. Geometrically,
the portfolio with maximum Sharpe ratio is the point where a line
through the origin is tangent to the efficient frontier, and therefore
it is also called the "tangency portfolio". 

In order to find the tangency point $(s_{TGP},\rho_{TGP})$ we observe
that the slope of the tangency line:
\begin{equation}
\frac{s_{TGP}-0}{\rho_{TGP}-0}=\frac{\sqrt{\frac{1}{d}(a_{11}\rho_{TGP}^{2}-2a_{12}\rho_{TGP}+a_{22})}}{\rho_{TGP}}
\end{equation}
should be equal with the derivative of the "efficient frontier"
at that point:
\begin{equation}
\frac{ds}{d\rho}=\frac{a_{11}\rho_{TGP}-a_{12}}{d\sqrt{\frac{1}{d}(a_{11}\rho_{TGP}^{2}-2a_{12}\rho_{TGP}+a_{22})}}.
\end{equation}
Thus, we easily obtain the risk-return pair for the "tangency-portfolio":
\begin{equation}
s_{TGP}=\sqrt{a_{22}}/a_{12},
\end{equation}
\begin{equation}
\rho_{TGP}=a_{22}/a_{12}.
\end{equation}
Also, the allocation of the assets for the "tangency portfolio"
are therefore given by:
\begin{equation}
\mathbf{w}_{TGP}=\mathbf{f}+\rho_{TGP}\mathbf{g}.
\end{equation}

\section{Eigen-portfolios}

The covariance matrix $\mathbf{S}$ is positive definite, and the
correlation matrix $\mathbf{C}$ is given by:
\begin{equation}
\mathbf{C}=\mathbf{\boldsymbol{\Omega}}^{-1}\mathbf{S}\mathbf{\boldsymbol{\Omega}}^{-1},
\end{equation}
where 
\begin{equation}
\boldsymbol{\Omega}=diag(\sqrt{s_{11}},\sqrt{s_{22}},...,\sqrt{s_{NN}})
\end{equation}
Obviously, the correlation matrix $\mathbf{C}$ has $N$ positive
eigenvalues:
\begin{equation}
\lambda_{1}\geq\lambda_{2}\geq...\geq\lambda_{N}\geq 0,
\end{equation}
and $N$ associated orthogonal eigenvectors:
\begin{equation}
\mathbf{V}=[\mathbf{v}^{(1)},\mathbf{v}^{(2)},...,\mathbf{v}^{(N)}],
\end{equation}
\begin{equation}
\left(\mathbf{v}^{(i)}\right)^{T}\mathbf{v}^{(j)}=\delta_{ij}=\left\{ \begin{array}{ccc}
1 & if & i=j\\
0 & if & i\neq j
\end{array}\right.,
\end{equation}
such that:
\begin{equation}
\mathbf{C}=\mathbf{V}\boldsymbol{\Lambda}\mathbf{V}^{T},
\end{equation}
where:
\begin{equation}
\boldsymbol{\Lambda}=diag(\lambda_{1},\lambda_{2},...,\lambda_{N}).
\end{equation}
In order to define the eigen-portfolios [7] we divide each eigenvector
of the correlation matrix by the volatility of the corresponding asset:
\begin{equation}
\mathbf{\boldsymbol{\xi}}^{(n)}=\mathbf{\boldsymbol{\Omega}}^{-1}\mathbf{v}^{(n)},\quad n=1,2,...,N,
\end{equation}
and we normalize by imposing a constant invested wealth, such that:
\begin{equation}
\boldsymbol{\psi}^{(n)}=\mathbf{\boldsymbol{\xi}}^{(n)}/\left(\mathbf{u}^{T}\mathbf{\boldsymbol{\xi}}^{(n)}\right)=\alpha_{n}^{-1}\mathbf{\boldsymbol{\xi}}^{(n)},\quad n=1,2,...,N,
\end{equation}
where $\mathbf{u}=[1,1,\ldots,1]^{T}$ , $\alpha_{n}=\mathbf{u}^{T}\mathbf{\boldsymbol{\xi}}^{(n)}$.
For each eigen-portfolio, the weight of a given asset is
inversely proportional to its volatility. Also, the eigen-portfolios
are pairwise orthogonal, and therefore completely decorrelated, since:
\begin{equation}
\left(\boldsymbol{\psi}^{(i)}\right)^{T}\mathbf{S}\boldsymbol{\psi}^{(j)}=\left\{ \begin{array}{ccc}
\alpha_{i}^{-1}\alpha_{j}^{-1}\lambda_{i} & if & i=j\\
0 & if & i\neq j
\end{array}\right..
\end{equation}
Thus, any portfolio can be represented as a linear combination of
the eigen-portfolios, since they are orthogonal and form a basis in
the asset space. 

It is also important to emphasize that the first eigen-portfolio,
corresponding to the largest eigenvalue, typically has positive weights,
corresponding to long-only positions. This is a consequence of the
classical Perron-Frobenius theorem, which states that a sufficient condition
for the existence of a dominant eigen-portfolio with positive entries
is that all the pairwise correlations are positive. One can always get a dominant eigen-portfolio with positive weights using a shrinkage estimate [8-9], which is a convex combination of the covariance matrix and a shrinkage target
matrix $\mathbf{D}$:
\begin{equation}
\tilde{\mathbf{S}}=(1-\gamma)\mathbf{S}+\gamma\mathbf{D},
\end{equation}
where the shrinkage matrix $\mathbf{D}$ is a diagonal matrix:
\begin{equation}
\mathbf{D}=diag(s_{11},s_{22},...,s_{NN}).
\end{equation}
In this case, there exists a $\gamma>0$ such that the shrinkage estimator
has a Dominant Eigen-Portfolio (DEP) with all weights positive. This portfolio 
is of interest since it provides a long-only investment solution, which may be 
desirable for investors who would like to avoid short positions and high risk.

\section{\textit{N} risky assets and a risk-free asset}

Let us now assume that one can also invest in a risk-free asset. A
risk-free asset $A_{f}$ is an asset with a low return $r_{f}$, but
with no risk at all, i.e. zero variance $s_{f}=0$. The risk-free
asset is also uncorrelated with the risky assets, such that $cov(A_{f},A_{n})=0$
for all risky assets $n=1,2,...,N$. The investor can both lend and
borrow at the risk-free rate. Lending means a positive amount is invested
in the risk-free asset, borrowing implies that a negative amount is
invested in the risk-free asset. In this case, we consider the following
quadratic optimization problem [1-3]:

\begin{equation}
\mathbf{w}=\arg\min_{\mathbf{w}}\left\{ \mathbf{w}^{T}\mathbf{S}\mathbf{w}\right\} ,
\end{equation}
subject to:
\begin{equation}
\mathbf{w}^{T}(\mathbf{r}-r_{f}\mathbf{u})+r_{f}=\rho.
\end{equation}
The Lagrangian of the problem is given by: 
\begin{equation}
L(\mathbf{w},\mu)=\mathbf{w}^{T}\mathbf{S}\mathbf{w}-\mu[\mathbf{w}^{T}(\mathbf{r}-r_{f}\mathbf{u})+r_{f}-\rho].
\end{equation}
The critical point of the Lagrangian is the solution of the system
of equations:
\begin{equation}
\nabla_{\mathbf{w}}L(\mathbf{w},\mu)=2\mathbf{Sw}-\mu(\mathbf{r}-r_{f}\mathbf{u})=0,
\end{equation}
\begin{equation}
\frac{\partial L(\mathbf{w},\mu)}{\partial\mu}=\mathbf{w}^{T}(\mathbf{r}-r_{f}\mathbf{u})+r_{f}-\rho=0,
\end{equation}
From the first equation we have:
\begin{equation}
\mathbf{w}=\frac{\mu}{2}\mathbf{S}^{-1}(\mathbf{r}-r_{f}\mathbf{u}).
\end{equation}
Therefore, the second equation becomes:
\begin{equation}
\frac{\mu}{2}(\mathbf{r}-r_{f}\mathbf{u})^{T}\mathbf{S}^{-1}(\mathbf{r}-r_{f}\mathbf{u})=\rho-r_{f},
\end{equation}
and from here we obtain:
\begin{equation}
\mu=2\frac{\rho-r_{f}}{b},
\end{equation}
where
\begin{equation}
b=a_{11}r_{f}^{2}-2a_{12}r_{f}+a_{22}.
\end{equation}
The weights of the risky assets are therefore given by:
\begin{equation}
\mathbf{w}(\rho)=\frac{\rho-r_{f}}{b}\mathbf{S}^{-1}(\mathbf{r}-r_{f}\mathbf{u}),
\end{equation}
and the corresponding amount that is invested in the risk-free asset
is:
\begin{equation}
w_{f}(\rho)=1-\mathbf{w}^{T}(\rho)\mathbf{u}=1-\frac{\rho-r_{f}}{b^{2}}(a_{12}-a_{11}r_{f}).
\end{equation}
Also, the standard deviation of the risky assets is:
\begin{equation}
s(\rho)=\sqrt{\mathbf{w}^{T}(\rho)\mathbf{S}\mathbf{w}(\rho)}=\frac{1}{\sqrt{b}}(\rho-r_{f}),
\end{equation}
or equivalently:
\begin{equation}
\rho(s)=r_{f}+\sqrt{b}s,
\end{equation}
This is the efficient frontier when the risk-free asset is added,
or the Capital Market Line (CML), and it is a straight
line in the return-risk ($\rho,s$) space. Obviously, CML intersects the
return axis for $s=0$, at $\rho=r_{f}$, which is the return when
the whole capital is invested in the risk-free asset. 

The tangency point of intersection between the efficient frontier
and the CML corresponds to the "market portfolio". This is the
portfolio on the CML where nothing is invested in the risk-free asset.
If the investor goes on the left side of the market portfolio, then
he invests a proportion in the risk-free asset. If he chooses the
right side of the market portfolio, he borrows at the risk-free rate. 

The market portfolio can be easily calculated from the equality condition:
\begin{equation}
s=\frac{\rho-r_{f}}{\sqrt{b}}=\sqrt{\frac{1}{d}(a_{11}\rho^{2}-2a_{12}\rho+a_{22})}.
\end{equation}
The solution of the above equation provides the coordinates of the
market portfolio:
\begin{equation}
s{}_{MP}=\frac{\sqrt{b}}{a_{12}-a_{11}r_{f}},
\end{equation}
\begin{equation}
\rho_{MP}=\frac{a_{22}-a_{12}r_{f}}{a_{12}-a_{11}r_{f}}.
\end{equation}
and the weights of the market portfolio are then given by: 
\begin{equation}
\begin{split}
& \mathbf{w}_{MP}=\frac{\rho_{MP}-r_{f}}{b^{2}}\mathbf{S}^{-1}(\mathbf{r}-r_{f}\mathbf{u}) \\
&\quad\quad =\frac{1}{a_{12}-a_{11}r_{f}}\mathbf{S}^{-1}(\mathbf{r}-r_{f}\mathbf{u}).
\end{split}
\end{equation}

\begin{algorithm}[h]
\caption{\textit{N} risky assets portfolio optimization}
\small{
$N$ \# number of risky assets (given)

$\mathbf{r}\leftarrow[r_{1},r_{2},\ldots,r_{N}]^{T}$ \# expected
returns (given)

$M$ \# number of portfolios computed on the frontier

$\rho_{max}$; maximum value of risk considered

$\mathbf{S}\leftarrow\mathrm{cov}(\mathbf{r})$

$\mathbf{Q}\leftarrow\mathbf{S}^{-1}$

$\mathbf{u}\leftarrow[1,1,\ldots,1]^{T}$

$\mathbf{A}\leftarrow\left[\begin{array}{cc}
a_{11} & a_{12}\\
a_{21} & a_{22}
\end{array}\right]\leftarrow\left[\begin{array}{cc}
\mathbf{u}^{T}\mathbf{Q}\mathbf{u} & \mathbf{r}^{T}\mathbf{Q}\mathbf{u}\\
\mathbf{r}^{T}\mathbf{Q}\mathbf{u} & \mathbf{r}^{T}\mathbf{Q}\mathbf{r}
\end{array}\right]$

$d\leftarrow a_{11}a_{22}-a_{12}^{2}$

$\mathbf{f}\leftarrow\frac{1}{d}\mathbf{Q}\left(a_{22}\mathbf{u}-a_{12}\mathbf{r}\right)$

$\mathbf{g}\leftarrow\frac{1}{d}\mathbf{Q}\left(-a_{12}\mathbf{u}+a_{11}\mathbf{r}\right)$

$\boldsymbol{\rho}\leftarrow[\rho_{1},\rho_{2},\ldots,\rho_{M}]^{T}$
\# return values on the frontier 

$\boldsymbol{s}\leftarrow[s_{1},s_{2},\ldots,s_{M}]^{T}$ \# risk
values on the frontier 

for($m=1$ to $m=M$)\{

$\quad$$\rho_{m}\leftarrow m\rho_{max}/M$

$\quad$$s_{m}\leftarrow\sqrt{\frac{a_{11}}{d}\left(\rho_{m}-\frac{a_{12}}{a_{11}}\right)^{2}+\frac{1}{a_{11}}}$

$\quad$\}

$\boldsymbol{W}\leftarrow[\mathbf{w}^{(1)},\mathbf{w}^{(2)},\ldots,\mathbf{w}^{(M)}]^{T}$
\# portfolio weights

for($m=1$ to $m=M$)\{

$\quad$$\mathbf{w}^{(m)}\leftarrow\mathbf{f}+\rho_{m}\mathbf{g}$

$\quad$\}

$s_{MVP}\leftarrow\sqrt{1/a_{11}}$ \# risk of MVP

$\rho_{MVP}\leftarrow\sqrt{a_{12}/a_{11}}$ \# return of MVP

$\mathbf{w}_{MVP}\leftarrow\mathbf{f}+\rho_{MVP}\mathbf{g}$ \# weights of MVP

$s_{TGP}\leftarrow\sqrt{a_{22}}/a_{12}$ \# risk of TGP

$\rho_{TGP}\leftarrow a_{22}/a_{12}$ \# return of TGP

$\mathbf{w}_{TGP}\leftarrow\mathbf{f}+\rho_{TGP}\mathbf{g}$ \# weights of TGP

$
\mathrm{return}\left(\mathbf{s},\boldsymbol{\rho},s_{MVP},\rho_{MVP},\mathbf{w}_{MVP},\rho_{TGP},s_{TGP},\mathbf{w}_{TGP},\mathbf{W}\right)
$
}
\end{algorithm}

\section{R implementation}

The code for portfolio optimization was written in R, which is a free software environment for statistical computing and graphics [10]. 
To exemplify the above analytical results, we consider a portfolio
of common stocks. The raw data can be downloaded from Yahoo finance [11], 
and contains historical prices of each stock. The list
of stocks to be extracted is given in a text file, as a comma delimited
list. The raw data corresponding to each stock is downloaded and saved
in a local "data" directory, using the "data.r" script (Appendix A), which
has one input argument: the file containing stock symbols.

Once the raw data is downloaded the correct daily closing prices for
each stock are extracted, and saved in another file, which is
the main data input for the optimization program. The extraction is
performed using the "price.r" script (Appendix B), which has three input arguments:
the file containing stock symbols included in the portfolio, the number
of trading days used in the model, the output file of the stock
prices.

The pseudo-code for the case with \textit{N} risky assets is presented
in Algorithm 1. Also, the R script performing the optimization and visualization for the \textit{N} risky assets case is "optimization1.r" (Appendix C). 
The script has three input arguments: the name of the data file, the number
of portfolios on the efficient frontier to be calculated, and the
maximum return considered on the "efficient frontier" (this should
be several (5-10) times higher than the maximum return of the individual
assets). 

The pseudo-code for the case with \textit{N} risky assets and a risk-free
asset is presented in Algorithm 2, and the R script performing the optimization and visualization for the
\textit{N} risky assets case is "optimization2.r" (Appendix D). The script
has four input arguments: the name of the data file, the number
of portfolios on the CML to be calculated, the daily return of the
risk free asset, and the maximum return considered on the "efficient
frontier". 

\begin{algorithm}
\small{
\caption{\textit{N} risky assets and a risk-free asset optimization}

$N$ \# number of risky assets

$\mathbf{r}\leftarrow[r_{1},r_{2},\ldots,r_{N}]^{T}$ \# expected
returns 

$r_{f}$; \# return of the risk free asset 

$M$ \# number of portfolios to be computed on the CML 

$\rho_{max}$; maximum value of risk considered

$\mathbf{S}\leftarrow\mathrm{cov}(\mathbf{r})$

$\mathbf{Q}\leftarrow\mathbf{S}^{-1}$

$\mathbf{u}\leftarrow[1,1,\ldots,1]^{T}$

$\mathbf{A}\leftarrow\left[\begin{array}{cc}
a_{11} & a_{12}\\
a_{21} & a_{22}
\end{array}\right]\leftarrow\left[\begin{array}{cc}
\mathbf{u}^{T}\mathbf{Q}\mathbf{u} & \mathbf{r}^{T}\mathbf{Q}\mathbf{u}\\
\mathbf{r}^{T}\mathbf{Q}\mathbf{u} & \mathbf{r}^{T}\mathbf{Q}\mathbf{r}
\end{array}\right]$

$b\leftarrow a_{11}r_{f}^{2}-2a_{12}r_{f}+a_{22}$

$\boldsymbol{\rho}\leftarrow[\rho_{1},\rho_{2},\ldots,\rho_{M}]^{T}$
\# return values on the CML 

$\boldsymbol{s}\leftarrow[s_{1},s_{2},\ldots,s_{M}]^{T}$ \# risk
values on the CML

for($m=1$ to $m=M$)\{

$\quad$$\rho_{m}\leftarrow m\rho_{max}/M$

$\quad$$s_{m}\leftarrow(\rho-r_{f})/\sqrt{b}$

$\quad$\}

$\boldsymbol{W}\leftarrow[\mathbf{w}^{(1)},\mathbf{w}^{(2)},\ldots,\mathbf{w}^{(M)}]^{T}$
\# portfolio weights on CML

$\boldsymbol{w}_{f}\leftarrow[w_{f1},w_{f2},\ldots,w_{fM}]^{T}$ \#
risk free asset weights on CML

for($m=1$ to $m=M$)\{

$\quad$$\mathbf{w}^{(m)}\leftarrow(\rho_{m}-r_{f})\mathbf{Q}(\mathbf{r}-r_{f}\mathbf{u})/b$

$\quad$$w_{fm}\leftarrow1-(\rho_{m}-r_{f})(a_{12}-a_{11}r_{f})/b^{2}$

$\quad$\}

$s_{MP}\leftarrow\sqrt{b}/(a_{12}-a_{11}r_{f})$ \# risk of MP

$\rho_{MP}\leftarrow(a_{22}-a_{12}r_{f})/(a_{12}-a_{11}r_{f})$ \#
return of MP

$\mathbf{w}_{MP}\leftarrow\mathbf{Q}(\mathbf{r}-r_{f}\mathbf{u})/(a_{12}-a_{11}r_{f})$
\# weights of MP

$\mathrm{return}\left(\mathbf{s},\boldsymbol{\rho},s_{MP},\rho_{MP},\mathbf{w}_{MP},\mathbf{W}\right)$
}
\end{algorithm}

In order to exectute the code, on Unix/Linux platforms one can simply run the following script:

\begin{lstlisting}[basicstyle={\footnotesize\ttfamily},breaklines=true,showstringspaces=false,tabsize=4]
# File containing the list of stock symbols
stocks="stocks.txt"
# Number of trading days
T=250
# File containing the daily stock prices
prices="portfolio.txt"
# Number of portfolios on the frontier
N=100
# Daily return for the risk free asset
R=0.0003
# Maximum daily return value considered
Rmax=0.01

./data.r $stocks
./price.r $stocks $T $prices
./optimization1.r $prices $N $Rmax
./optimization2.r $prices $N $R $Rmax
\end{lstlisting}

On Windows platforms one can use a simple batch file, like the following ones:

\begin{lstlisting}[basicstyle={\footnotesize\ttfamily},breaklines=true,showstringspaces=false,tabsize=4]
@ECHO OFF
REM File containing the list of stock symbols
SET stocks="stocks.txt"
REM Number of trading days
SET T="250"
REM File containing the daily stock prices
SET file="portfolio.txt"
REM Number of portfolios on the frontier 
SET N="100" 
REM Daily return for the risk free asset
SET R="0.0003"
REM Maximum daily return value considered 
SET Rmax="0.01"
REM The path to R
SET rpath="C:\Program Files\R\R-3.0.1\bin\x64\Rscript.exe"

CALL %rpath% data.r %stocks%
CALL %rpath% price.r %stocks% %T% %file%
CALL %rpath% optimization1.r %file% %N% %Rmax%
CALL %rpath% optimization2.r %file% %N% %R% %Rmax%
\end{lstlisting}

In these examples: "stocks.txt" is a file containing the symbols
of some common stocks to be downloaded; "portfolio.txt"
is the file where all the relevant stock prices are extracted for
the current analysis.

\section{Numerical examples}

In order to illustrate the above results, we consider the case of
a portfolio consisting of $N=10$ common stocks from IT industry.
The content of the "stocks.txt" file is:
\begin{equation}
\begin{split}
& \mathrm{FB,INT,AAPL,MSFT,ORCL,} \\
& \mathrm{GOOG,YHOO,DELL,IBM,HPQ}
\end{split}
\end{equation}
A historical record of daily prices of these stocks for the last $T=250$
trading days was used to estimate the mean return and the covariance
matrix. The maximum return considered in computation is 0.01. 

The daily returns of the assets are calculated as: 
\begin{equation}
R(n,t)=\frac{p(n,t+1)-p(n,t)}{p(n,t)},
\end{equation}
where $t=1,2,...,T-1$ is the day index, and $p(n,t)$ is the price
of asset $A_{n}$ at the closing day $t$. The estimate average returns
and covariances are:
\begin{equation}
r_{n}=\frac{1}{T-1}\sum_{t=1}^{T-1}R(n,t),
\end{equation}
\begin{equation}
s_{ij}=\frac{1}{T-1}\sum_{t=1}^{T-1}[R(i,t)-r_{i}][R(j,t)-r_{j}],
\end{equation}
$n,i,j=1,2,...,N$

The asset prices and their expected returns for the considered time period are given in Figure 1. The resulted efficient frontier is given in Figure 2. The figure shows  also the risk-return values of each stock considered, the minimum variance portfolio MVP1, 
the tangency portfolio TGP, and the weights of the efficient frontier portfolios as a function of risk. Figure 3 provides the weights of the MVP1, TGP and DEP portfolios.

In order to illustrate the effect of the risk free asset (RFA), we
consider that the investor can invest at a daily rate of return of
$r_{f}=0.0003$. The capital market line, together with the frontier
line and the position of the new minimum variance portfolio MVP2 and
of the market portfolio MP are given in Figure 4. In this figure we also 
the risk dependence of the optimal portfolios from the capital market
line, and we emphasize the MVP2 and MP portfolios. The weights of
MVP2 and MP are plotted in Figure 5. 

\begin{figure}
\centering
\includegraphics[scale=0.7]{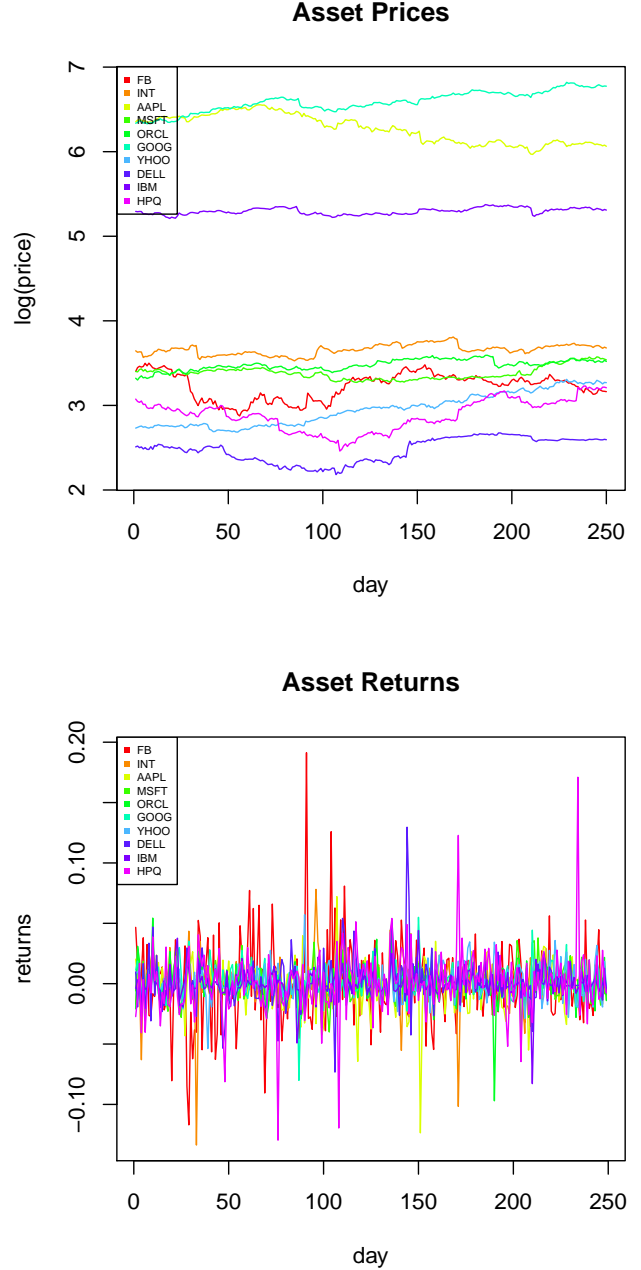}
\caption{Asset prices (log scale) and their daily returns.}
\end{figure}

\begin{figure}
\centering
\includegraphics[scale=0.7]{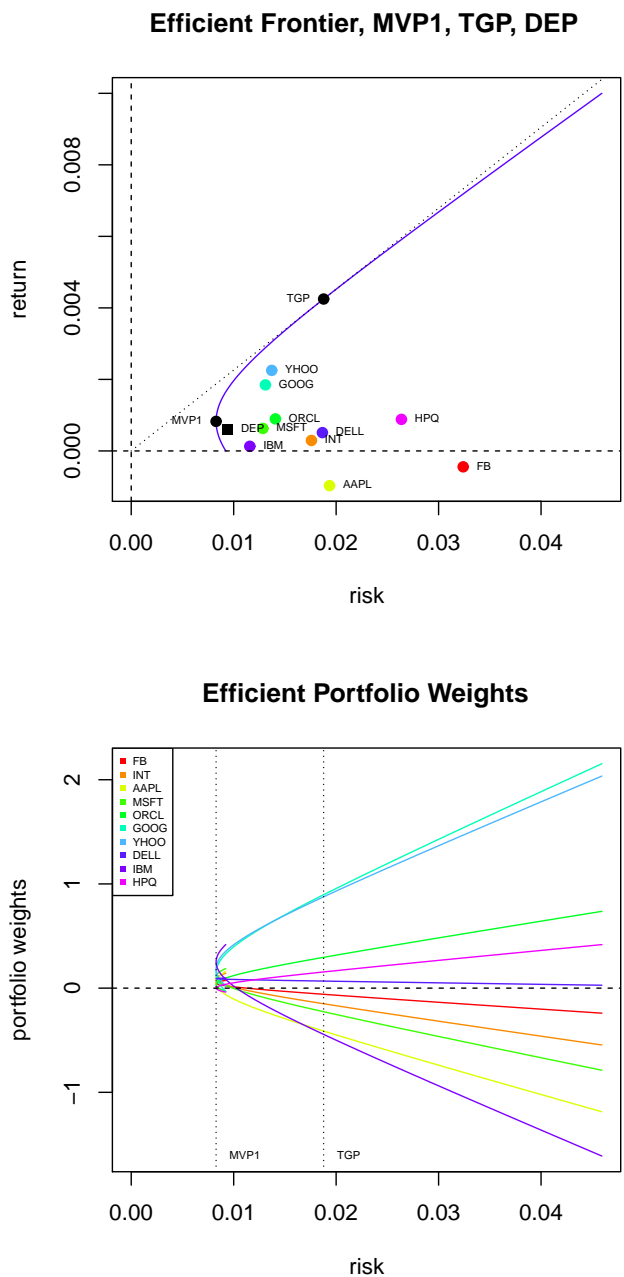}
\caption{Efficient frontier and risk dependence of portfolio weights.}
\end{figure}

\begin{figure}
\centering
\includegraphics[scale=0.7]{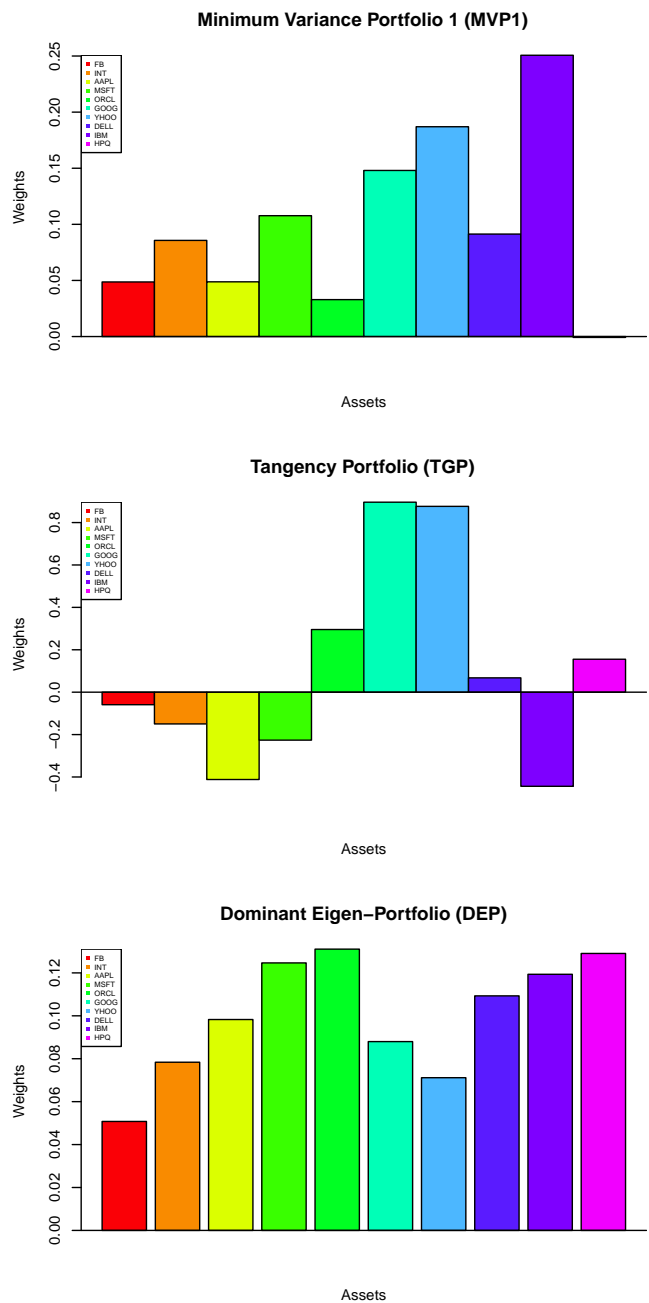}
\caption{Weights of MVP1, TGP and DEP portfolios.}
\end{figure}

\begin{figure}
\centering
\includegraphics[scale=0.7]{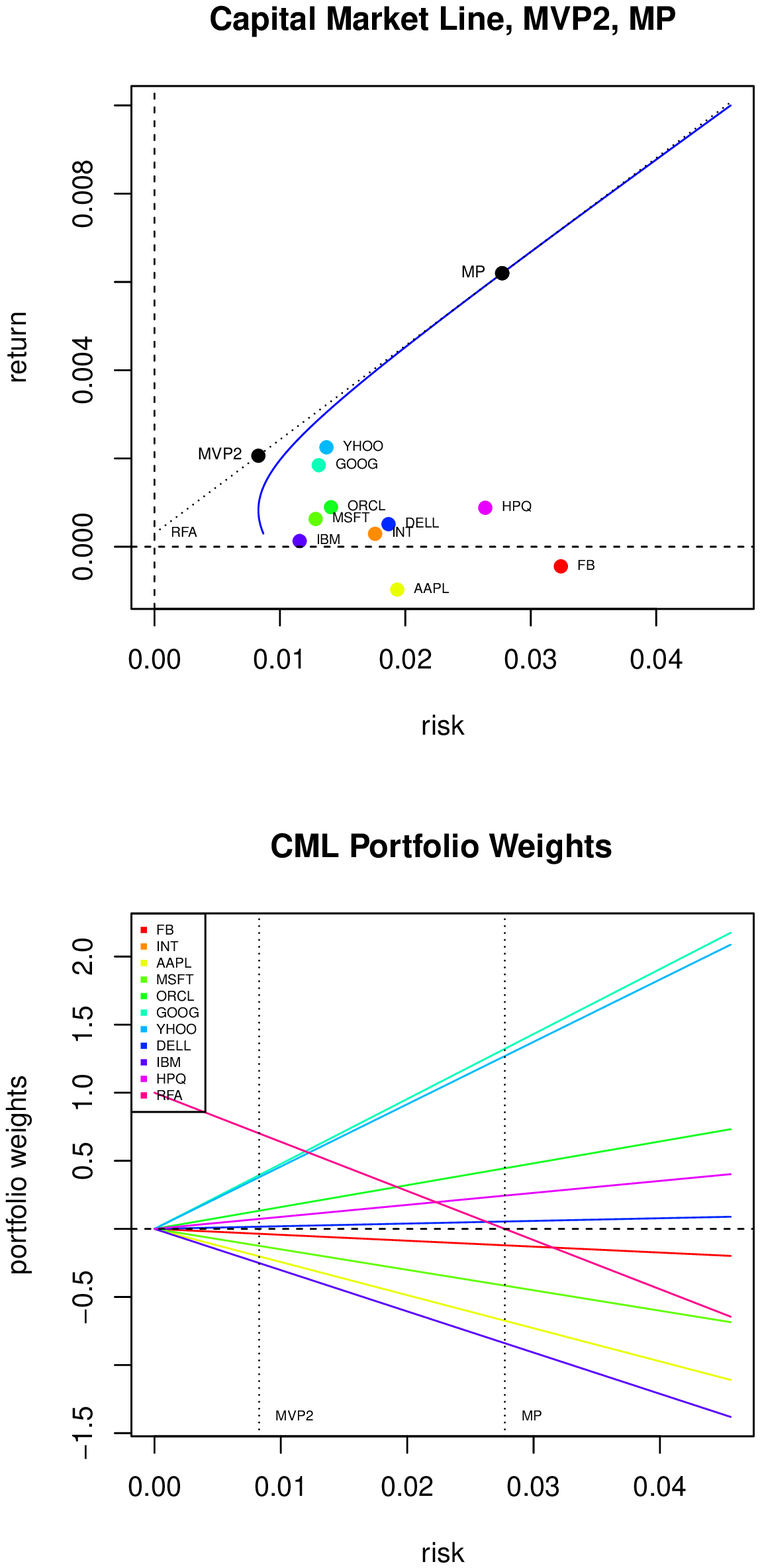}
\caption{Capital market line and risk dependence of portfolio weights.}
\end{figure}

\begin{figure}
\centering
\includegraphics[scale=0.7]{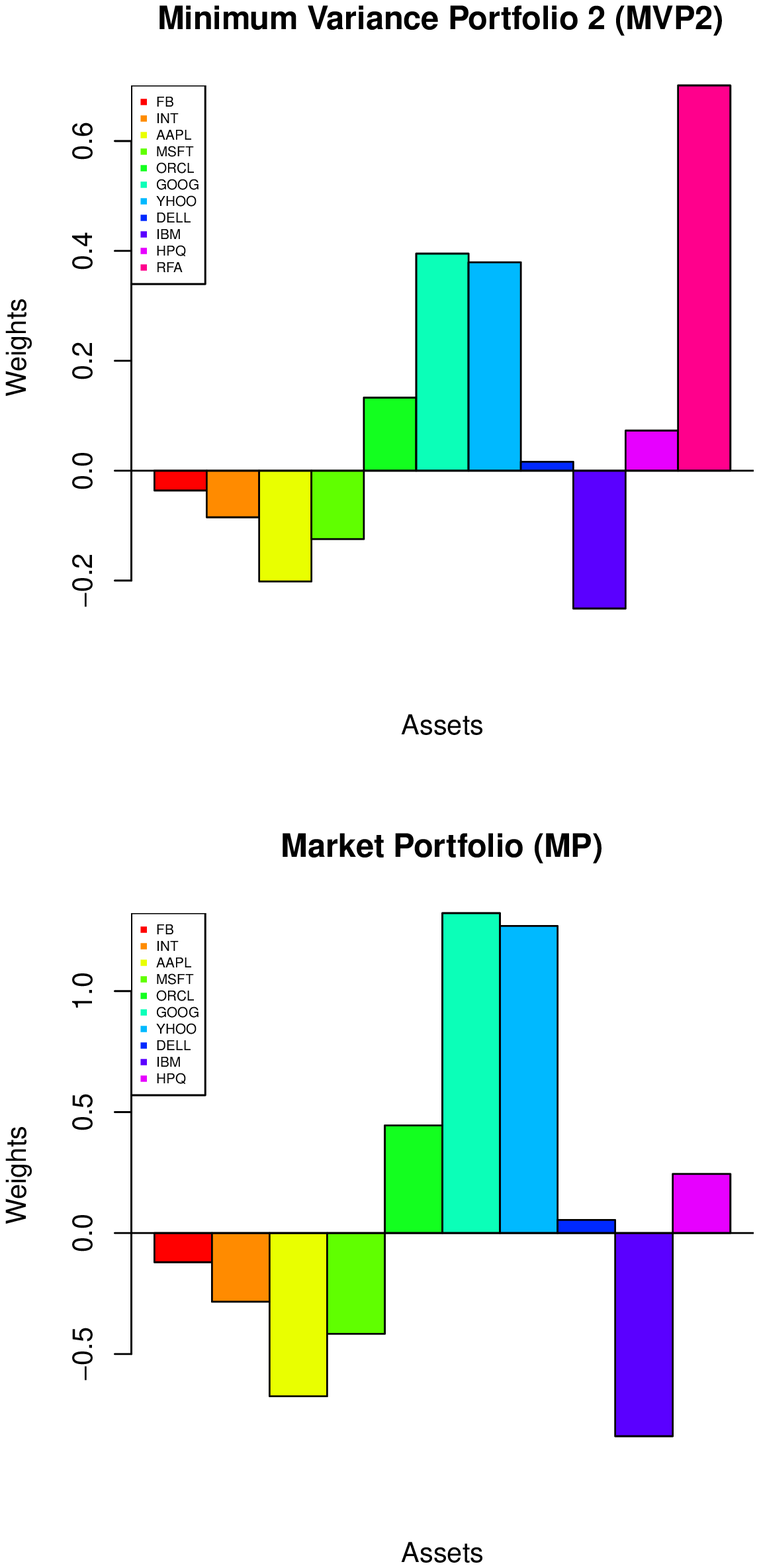}
\caption{Weights of MVP2 and MP portfolios.}
\end{figure}

\section{Conclusion}
We have considered the standard risk-return portfolio optimization
model, when both long buying and short selling, i.e. positive and
negative weights, of a relatively large number of assets is allowed.
We have derived the analytical expression of the efficient frontier
for a portfolio of $N$ risky assets, and of the capital market line
when a risk-free asset is added to the model. Also, we have provided
an R implementation for both cases, and we have discussed in detail
a numerical example of a portfolio of several risky common stocks. 

\appendices
\section{}
\begin{lstlisting}[basicstyle={\scriptsize\ttfamily},breaklines=true,showstringspaces=false,tabsize=4]
#!/usr/bin/Rscript
# data.r, downloads raw stock data

args <- commandArgs(TRUE)

extract <- function(args){
stocks <- t(read.table(args[1],sep=",")[1,])
J <- length(stocks)
dir.create("data", showWarnings = FALSE)
path <- "http://ichart.finance.yahoo.com/table.csv?s="
for(j in 1:J){
	dat <- read.csv(paste(path,stocks[j],sep=""))
	write.csv(dat, file=paste("./data/",stocks[j],sep=""),
			row.names=FALSE, quote=FALSE)
	}
}

extract(args)
\end{lstlisting}

\section{}
\begin{lstlisting}[basicstyle={\scriptsize\ttfamily},breaklines=true,showstringspaces=false,tabsize=4]
#!/usr/bin/Rscript
# price.r, extracts the stock prices

args <- commandArgs(TRUE)

extract_price <- function(args){
stocks <- t(read.table(args[1], sep=",")[1,])	
J <- length(stocks)
N <- as.integer(args[2])	
dat <- read.csv(paste("./data/", stocks[1], sep=""))
price <- dat[1:N,5]
if(J > 1){
	for(j in 2:J){
		dat <- read.csv(paste("./data/", 
				stocks[j], sep=""))
		price <- cbind(price, dat[1:N,5])
		}
	}
write.table(price, file=paste("./data/", args[3], sep=""),
		row.names=FALSE, col.names=stocks, 
		quote=FALSE, sep=",")
}

extract_price(args)
\end{lstlisting}

\section{}
\begin{lstlisting}[basicstyle={\scriptsize\ttfamily},breaklines=true,showstringspaces=false,tabsize=4]
#!/usr/bin/Rscript
# optimization1.r, N risky assets case

args <- commandArgs(TRUE)

read_data <- function(args){
data <- read.table(paste("./data/", args[1], sep=""),
		header=TRUE, sep=",");
return(list(data, as.integer(args[2]), 
		as.double(args[3]))); 
}

returns <- function(data){
dat <- data[[1]]; N <- nrow(dat) - 1; J <- ncol(dat);
ret <- (dat[1:N,1] - dat[2:(N+1),1])/dat[2:(N+1),1];
if(J > 1){
	for(j in 2:J){
		ret <- cbind(ret, (dat[1:N,j] - dat[2:(N+1),j])/
				dat[2:(N+1),j])
	}
}
return(list(ret, names(data[[1]]), data[[2]], data[[3]]));
}

optimization <- function(returns){
p <- colMeans(returns[[1]]); names(p) <- returns[[2]]; 
J <- ncol(returns[[1]]); M <- returns[[3]];
Rmax <- returns[[4]];
S <- cov(returns[[1]]); Q <- solve(S); u <- rep(1,J);
a <- matrix(rep(0,4),nrow=2);
a[1,1] <- u%*%Q%*%u;
a[1,2] <- a[2,1] <- u%*%Q%*%p;
a[2,2] <- p%*%Q%*%p;
d <- a[1,1]*a[2,2] - a[1,2]*a[1,2];
f <- (Q%*%( a[2,2]*u - a[1,2]*p))/d;
g <- (Q%*%(-a[1,2]*u + a[1,1]*p))/d;
r <- seq(0, Rmax, length=M);
w <- matrix(rep(0,J*M), nrow=J);
for(m in 1:M) w[,m] <- f + r[m]*g;
s <- sqrt( a[1,1]*((r - a[1,2]/a[1,1])^2)/d + 1/a[1,1]);
ss <- sqrt(diag(S));
minp <- c(sqrt(1/a[1,1]), a[1,2]/a[1,1]);
wminp <- f + (a[1,2]/a[1,1])*g;
tanp <- c(sqrt(a[2,2])/a[1,2], a[2,2]/a[1,2]);
wtanp <- f + (a[2,2]/a[1,2])*g;	
Q <- sqrt(diag(1.0/ss)); 
x <- eigen(Q%*%S%*%Q);
v <- Q%*%x$vec;
for(j in 1:J) v[,j] <- v[,j]/(u%*%v[,j]);
sv <- rv <- rep(0, J);
for(j in 1:J){
	rv[j] <- t(v[,j])%*%p;
	if(rv[j] < 0){
		rv[j] <- -rv[j];
		v[,j] <- -v[,j];
		}
	sv[j] <- sqrt(t(v[,j])%*%S%*%v[,j]);
	}
return(list(s, r, ss, p, minp, tanp, wminp, wtanp, 
		w, v, sv, rv));
}

plot_results<- function(data, returns, results){
dat <- log(data[[1]]); M <- nrow(dat); J <- ncol(dat);
ymax = max(dat); ymin = min(dat)
mycolors <- rainbow(J+1);
s <- results[[1]]; r <- results[[2]];
ss <- results[[3]];	p <- results[[4]];
minp <- results[[5]]; tanp <- results[[6]];
wminp <- results[[7]]; wtanp <- results[[8]];
f <- t(results[[9]]); v <- results[[10]];
sv <- results[[11]]; rv <- results[[12]];
postscript(file="./results1/fig1.eps", onefile=FALSE, 
			horizontal=FALSE, height=10, width=5);
par(mfrow=c(2,1));
id <- c(1:nrow(dat));
plot(id, rev(dat[,1]), ylim=c(ymin, ymax), type="l",
		col=mycolors[1], xlab="day", ylab="log(price)",
		main = "Asset Prices");
if(J > 1){
	for(j in 2:J){
		lines(id, rev(dat[,j]), type="l", 
				col=mycolors[j]);
	}
}
legend("topleft", names(dat), cex=0.5, pch=rep(15, J), 
		col=mycolors);
ret <- returns[[1]];
ymax = max(ret); ymin = min(ret);
id <- c(1:nrow(ret));
plot(id, rev(ret[,1]), ylim=c(ymin, ymax), type="l",
		col=mycolors[1], xlab="day", ylab="returns",
		main = "Asset Returns");
if(J > 1){
	for(j in 2:J){
		lines(id, rev(ret[,j]),type="l",col=mycolors[j]);
	}
}
legend("topleft", returns[[2]], cex=0.5, pch=rep(15, J), col=mycolors);
postscript(file="./results1/fig2.eps", onefile=FALSE, 
			horizontal=FALSE, height=10, width=5);
par(mfrow=c(2,1));
plot(s, r, xlim=c(0,max(s)), ylim=c(min(r,p), max(r,p)),
		type="l", col="blue", xlab="risk", ylab="return",
		main = "Efficient Frontier, MVP1, TGP");
points(ss, p, pch=19, col=mycolors);
text(ss, p, pos=4, cex=0.5, names(p));
points(sv[1], rv[1], pch=15, col="black");
text(sv[1], rv[1], pos=4, cex=0.5, "DEP");
points(minp[1], minp[2], pch=19, col="black");
text(minp[1], minp[2], pos=2, cex=0.5, "MVP1");
points(tanp[1], tanp[2], pch=19, col="black");
text(tanp[1], tanp[2], pos=2, cex=0.5, "TGP");
lines(c(0,max(s)), c(0,max(s)*tanp[2]/tanp[1]), lty=3);
abline(h=0, lty=2); abline(v=0, lty=2);
plot(s, f[,1], xlim=c(0,max(s)), ylim=c(min(f),max(f)),
		col=mycolors[1], type="l",
		xlab="risk", ylab="portfolio weights",
		main = "Efficient Portfolio Weights");
if(J > 1){
	for(j in 2:J){
		lines(s, f[,j], type="l", col=mycolors[j]);
	}
}
abline(h=0, lty=2); abline(v=minp[1], lty=3); 
abline(v=tanp[1], lty=3);
text(minp[1], min(f), pos=4, cex=0.5, "MVP1");
text(tanp[1], min(f), pos=4, cex=0.5, "TGP");
legend("topleft", names(p), cex=0.5, pch=rep(15, J), 
		col=mycolors);
postscript(file="./results1/fig3.eps", onefile=FALSE, 
			horizontal=FALSE, height=10, width=5);
par(mfrow=c(2,1));
barplot(wminp, main="Minimum Variance Portfolio 1 (MVP1)",
		xlab="Assets", ylab="Weights", 
		col=mycolors, beside=TRUE);
abline(h=0, lty=1);
legend("topleft", names(p), cex=0.5, pch=rep(15, J), 
		col=mycolors);
barplot(wtanp, main="Tangency Portfolio (TGP)",
		xlab="Assets", ylab="Weights", col=mycolors, 
		beside=TRUE);
abline(h=0, lty=1);
legend("topleft", names(p), cex=0.5, pch=rep(15, J), 
		col=mycolors);
barplot(v[,1], main="Dominant Eigen-Portfolio (DEP)",
		xlab="Assets", ylab="Weights", col=mycolors, beside=TRUE);
abline(h=0, lty=1);
legend("topleft", names(p), cex=0.5, pch=rep(15, J), col=mycolors);
}

data <- read_data(args);
returns <- returns(data);
dir.create("results1", showWarnings = FALSE);
results <- optimization(returns);
plot_results(data, returns, results);
\end{lstlisting}

\section{}
\begin{lstlisting}[basicstyle={\scriptsize \ttfamily},breaklines=true,showstringspaces=false,tabsize=4]
args <- commandArgs(TRUE);

read_data <- function(args){
data <- read.table(paste("./data/", args[1], sep=""),
		header=TRUE, sep=",")
return(list(data, as.integer(args[2]), as.double(args[3]), 
		as.double(args[4])));
}

returns <- function(data){
dat <- data[[1]]; N <- nrow(dat) - 1; J <- ncol(dat);
ret <- (dat[1:N,1] - dat[2:(N+1),1])/dat[2:(N+1),1];
if(J > 1){
	for(j in 2:J){
		ret <- cbind(ret, (dat[1:N,j] - dat[2:(N+1),j])/
				dat[2:(N+1),j]);
	}
}
return(list(ret, names(data[[1]]), data[[2]], 
		data[[3]], data[[4]]));
}

foptimization <- function(returns){
p <- colMeans(returns[[1]]); names(p) <- returns[[2]];
J <- ncol(returns[[1]]); M <- returns[[3]]; 
R <- returns[[4]]; Rmax <- returns[[5]];
S <- cov(returns[[1]]); Q <- solve(S); u <- rep(1,J); 
a <- matrix(rep(0,4),nrow=2);
a[1,1] <- u%*%Q%*%u;
a[1,2] <- a[2,1] <- u%*%Q%*%p;
a[2,2] <- p%*%Q%*%p;
d <- a[1,1]*a[2,2] - a[1,2]*a[1,2];
r <- seq(R, Rmax, length=M);
s <- sqrt( a[1,1]*((r - a[1,2]/a[1,1])^2)/d + 1/a[1,1]);
ss <- sqrt(diag(S));
cml <- c(sqrt(a[1,1]*R*R - 2*a[1,2]*R + a[2,2]), R);
z <- (r - R)/cml[1]; f <- Q%*%(p - R*u)/(cml[1]*cml[1]);
wcml <- matrix(rep(0,J*M), nrow=J); wf <- rep(0,M);
for(m in 1:M){
	wcml[,m] <- (r[m] - R)*f; wf[m] <- 1 - wcml[,m]%*%u;
	}
wcml <- rbind(wcml, t(wf));
mp <- c(cml[1]/(a[1,2] - a[1,1]*R),
		(a[2,2] - a[1,2]*R)/(a[1,2] - a[1,1]*R));
wmp <- Q%*%(p - R*u)/(a[1,2] - a[1,1]*R);
minp <- c(sqrt(1/a[1,1]), cml[1]*sqrt(1/a[1,1]) + R);
wminp <- (minp[2] - R)*f; wfminp <- 1- t(wminp)%*%u;
wminp <- rbind(wminp, wfminp);
return(list(s, z, r, ss, p, cml, wcml, mp, wmp, 
		minp, wminp));
}

plot_results<- function(data, returns, results){
dat <- log(data[[1]]); M <- nrow(dat); J <- ncol(dat);
ymax = max(dat); ymin = min(dat); mycolors <- rainbow(J);
s <- results[[1]]; z <- results[[2]]; r <- results[[3]]; 
ss <- results[[4]]; p <- results[[5]]; 
cml <- results[[6]];
mp <- results[[8]]; wmp <- results[[9]];
minp <- results[[10]]; wminp <- results[[11]];
postscript(file="./results2/fig1.eps", onefile=FALSE, 
		horizontal=FALSE, height=10, width=5);
par(mfrow=c(2,1));
id <- c(1:nrow(dat));
plot(id, rev(dat[,1]), ylim=c(ymin, ymax), type="l",
		col=mycolors[1], xlab="day", ylab="log(price)",
		main = "Asset Prices");
if(J > 1){
	for(j in 2:J){
		lines(id, rev(dat[,j]),type="l",col=mycolors[j]);
	}
}
legend("topleft", names(dat), cex=0.5, pch=rep(15, J), 
		col=mycolors);
ret <- returns[[1]]; ymax = max(ret); ymin = min(ret);
id <- c(1:nrow(ret));
plot(id,rev(ret[,1]),ylim=c(ymin, ymax),type="l",
		col=mycolors[1], xlab="day", ylab="returns",
		main = "Asset Returns");
if(J > 1){
	for(j in 2:J){
		lines(id, rev(ret[,j]),type="l",col=mycolors[j])
	}
}
legend("topleft", returns[[2]], cex=0.5, pch=rep(15, J), 
		col=mycolors);
postscript(file="./results2/fig2.eps", onefile=FALSE, 
		horizontal=FALSE, height=10, width=5);
par(mfrow=c(2,1));
mycolors <- rainbow(length(p)+1);
plot(s,r,xlim=c(0, max(s)),ylim=c(min(r, p),max(r, p)),
		type="l", col="blue", xlab="risk", ylab="return",
		main = "Capital Market Line, MVP2, MP");
points(ss, p, pch=19, col=mycolors);
text(ss, p, pos=4, cex=0.5, names(p));
points(mp[1], mp[2], pch=19, col="black");
points(mp[1], mp[2], pch=19, col="black");
text(mp[1], mp[2], pos=2, cex=0.6, "MP");
points(minp[1], minp[2], pch=19, col="black");
text(minp[1], minp[2], pos=2, cex=0.6, "MVP2");
text(0, cml[2], pos=4, cex=0.5, "RFA");
lines(c(0, max(s)), c(cml[2], max(s)*cml[1] + cml[2]), 
		lty=3);
abline(h=0, lty=2); abline(v=0, lty=2); 
f <- t(results[[7]]); mycolors <- rainbow(J+1);
plot(z,f[,1], xlim=c(0,max(z)), ylim=c(min(f),max(f)), 
		type="l", col=mycolors[1], xlab="risk", 
		ylab="portfolio weights", 
		main="CML Portfolio Weights");
if(J > 1){
	for(j in 2:J+1){
		lines(z,f[,j],type="l",col=mycolors[j]);
	}
}
abline(h=0, lty=2); abline(v=mp[1], lty=3); 
text(mp[1], min(f), pos=4, cex=0.5, "MP");
abline(v=minp[1], lty=3); 
text(minp[1], min(f), pos=4, cex=0.5, "MVP2");
legend("topleft", c(names(p), "RFA"),
		cex=0.5, pch=rep(15, J+1), col=mycolors);
postscript(file="./results2/fig3.eps", onefile=FALSE, 
		horizontal=FALSE, height=10, width=5);
par(mfrow=c(2,1));
barplot(wminp, main="Minimum Variance Portfolio 2",
		xlab="Assets", ylab="Weights", 
		col=mycolors, beside=TRUE);
abline(h=0, lty=1);
legend("topleft", c(names(p),"RFA"), cex=0.5, 
		pch=rep(15, J+1), col=mycolors);
barplot(wmp, main="Market Portfolio",
		xlab="Assets", ylab="Weights", 
		col=mycolors, beside=TRUE);
abline(h=0, lty=1);
legend("topleft", names(p), cex=0.5, 
		pch=rep(15, J), col=mycolors);
}

data <- read_data(args);
returns <- returns(data);
dir.create("results2", showWarnings = FALSE);
results <- foptimization(returns);
plot_results(data, returns, results);

\end{lstlisting}

\end{document}